\documentclass[apjl]{emulateapj}
\usepackage{amsmath,graphicx,epsfig,multirow,natbib,rotating,tabularx}
\usepackage[usenames]{color}
\newcommand{\longname}{HSC J142449$-$005322}
\newcommand{\horus}{{\it Eye of Horus}}
\newcommand{\zl}{0.795}
\newcommand{\zsone}{1.302}
\newcommand{\zstwo}{1.988}

\shorttitle{Double Source Plane Lens in HSC}
\shortauthors{Tanaka}

\begin{document}

\title{
  A Spectroscopically Confirmed Double Source Plane Lens System\\
  in the Hyper Suprime-Cam Subaru Strategic Program
}

\author{
Masayuki Tanaka\altaffilmark{1},
Kenneth Wong\altaffilmark{1},
Anupreeta More\altaffilmark{2},
Arsha Dezuka\altaffilmark{3},
Eiichi Egami\altaffilmark{4},
Masamune Oguri\altaffilmark{2,5,6},
Sherry H.~Suyu\altaffilmark{7,8},
Alessandro Sonnenfeld\altaffilmark{2},
Ryou Higuchi\altaffilmark{9},
Yutaka Komiyama\altaffilmark{1},
Satoshi Miyazaki\altaffilmark{1,10},
Masafusa Onoue\altaffilmark{10,1},
Shuri Oyamada\altaffilmark{11},
Yousuke Utsumi\altaffilmark{12}
}

\altaffiltext{1}{National Astronomical Observatory of Japan, Osawa 2-21-1, Mitaka, Tokyo 181-8588, Japan}
\altaffiltext{2}{Kavli Institute for the Physics and Mathematics of the Universe (Kavli IPMU, WPI), University of Tokyo, Chiba 277-8583, Japan}
\altaffiltext{3}{Department of Astronomy, University of Kyoto, Kitashirakawa-Oiwake-cho, Sakyo-ku, Kyoto 606-8502, Japan}
\altaffiltext{4}{Steward Observatory, University of Arizona, 933 North Cherry Avenue, Tucson, AZ 85721, USA}
\altaffiltext{5}{Department of Physics, University of Tokyo, 7-3-1 Hongo, Bunkyo-ku, Tokyo 113-0033, Japan}
\altaffiltext{6}{Research Center for the Early Universe, University of Tokyo, Tokyo 113-0033, Japan}
\altaffiltext{7}{Max Planck Institute for Astrophysics, Karl-Schwarzschild-Strasse 1, D-85740 Garching, Germany}
\altaffiltext{8}{Institute of Astronomy and Astrophysics, Academia Sinica, P.O.~Box 23-141, Taipei 10617, Taiwan}
\altaffiltext{9}{Institute for Cosmic Ray Research, The University of Tokyo, 5-1-5 Kashiwa-no-Ha, Kashiwa City, Chiba, 277-8582, Japan}
\altaffiltext{10}{SOKENDAI (The Graduate University for Advanced Studies), Mitaka, Tokyo 181-8588, Japan}
\altaffiltext{11}{Japan Women's University, 2-8-1 Mejirodai, Bunkyo, Tokyo 112-8681, Japan}
\altaffiltext{12}{Hiroshima Astrophysical Science Center, Hiroshima University, 1-3-1, Kagamiyama, Higashi-Hiroshima, Hiroshima 739-8526, Japan}

\begin{abstract}
We report the serendipitous discovery of \longname, a double source plane lens system in
the Hyper Suprime-Cam Subaru Strategic Program.  We dub the system \horus.
The lens galaxy is a very massive early-type galaxy with stellar mass of $\sim7\times10^{11}\rm\ M_\odot$
located at $z_{\mathrm{L}}=\zl$.  The system exhibits two arcs/rings with clearly different colors,
including several knots.  We have performed spectroscopic follow-up observations of
the system with FIRE on Magellan. The outer ring is confirmed at $z_{\mathrm{S}2}=\zstwo$ with multiple
emission lines, while the inner arc and counterimage is confirmed at $z_{\mathrm{S}1}=\zsone$.
This makes it the first double source plane system with spectroscopic redshifts of both sources.
Interestingly, redshifts of two of the knots embedded in the outer ring are
found to be offset by $\Delta z=0.002$ from the other knots, suggesting that the outer ring
consists of at least two distinct components in the source plane.
We perform lens modeling with two independent codes and
successfully reproduce the main features of the system.  However, two of the lensed sources
separated by $\sim0.7$ arcsec cannot be reproduced by a smooth potential, and 
the addition of substructure to the lens potential is required to reproduce them.
Higher-resolution imaging of the system will help decipher the origin of this lensing
feature and potentially detect the substructure.
\end{abstract}


\section{Introduction}
Strong gravitational lensing is a powerful tool to map the matter distribution around galaxies on small scales.  In particular, it is possible to infer the total mass enclosed within the Einstein radius, allowing us to separate the dark and baryonic mass distribution of massive galaxies and constrain the stellar initial mass function \citep[IMF; e.g.,][]{treu2010}.
However, these measurements are complicated by various model degeneracies.

Lenses with multiple sources at distinct redshifts offer a unique opportunity to break these degeneracies, leading to accurate measurements of the dark matter distribution and IMF \citep[e.g.,][]{sonnenfeld2012}.  In addition, such double source plane (DSP) lens systems can be a useful, complementary cosmological probe \citep[e.g.,][]{collett2012,collett2014,linder16}, provided the mass-sheet degeneracy is broken \citep[e.g.,][]{falco1985,schneider2014}.  However, DSP lenses are extremely rare.  Only a handful are currently known \citep[e.g.,][]{gavazzi2008,tu2009,cooray11,brammer12}, and spectroscopic redshifts of the second sources have not been measured for these systems due to their faintness.
Ongoing/upcoming wide-area imaging surveys are expected to increase
the number of such rare but fruitful systems.  Here, we report the discovery of
a DSP lens system from the Hyper Suprime-Cam Subaru Strategic Program (hereafter the HSC survey).

The HSC survey is the largest observing program ever approved at
the Subaru 8.2-m telescope (300 nights).  With a 3-layered survey strategy over
$\sim$1400 deg$^{2}$, it aims to address several outstanding
astrophysical questions such as the nature of dark energy, galaxy
evolution, and cosmic reionization.  The survey is $\sim20\%$ complete
as of this writing.  While the survey is still in an early phase, the exquisite
data quality has led to the exciting discovery of this DSP lens.
The system, \longname, was serendipitously discovered at
a science school for undergraduate students hosted at the National Astronomical
Observatory of Japan in 2015.  While visually inspecting galaxy cluster candidates,
we found a likely cluster member with a clear sign of gravitational lensing.
A more careful inspection revealed the multi-source nature of the system.  
We dub it \horus, as the system resembles it.

We describe our photometric and spectroscopic data in Section~\ref{sec:data}.  We discuss properties of the lens galaxy in Section~\ref{sec:lensgal} and describe our lens modeling procedures in Section~\ref{sec:lensmod}.  Our main results are presented in Section~\ref{sec:summary}.  We adopt $\rm H_0=70\ km\ s^{-1}\ Mpc^{-1}$,
$\Omega_M=0.27$ and $\Omega_\Lambda=0.73$.  Magnitudes are AB magnitudes.


\section{Data}  \label{sec:data}
\begin{figure*}
\begin{centering}
\plotone{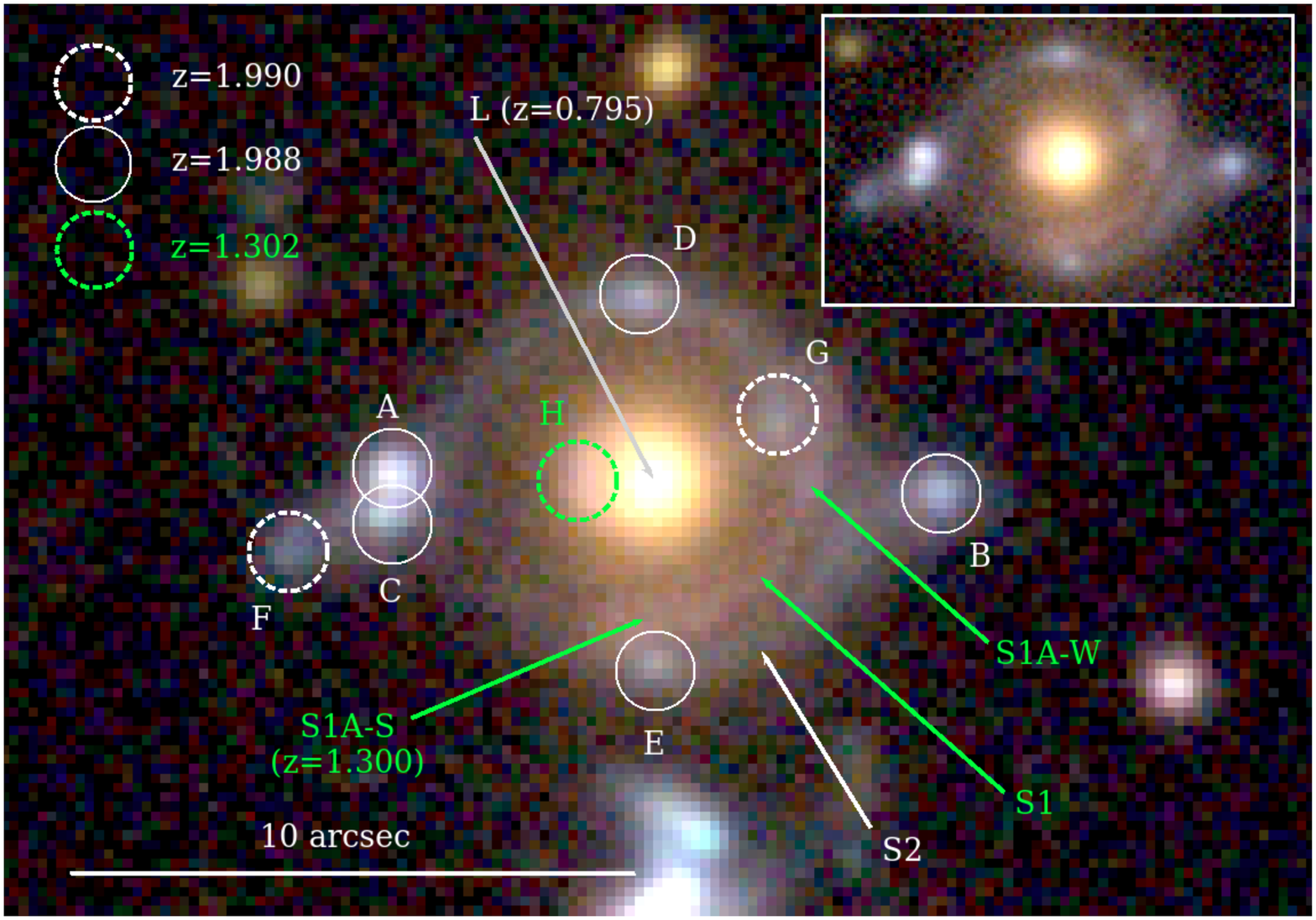}
\caption{
  $riz$ color composite of \horus.  The lens galaxy is labeled ``L", and the inner and outer sources are
  labeled ``S1" and ``S2", respectively.  The knots in the sources are labeled alphabetically in approximate
  order of brightness.  Note that the two rings clearly have different colors.  The spectroscopic redshifts of
  the various features are indicated by the colored circles in the upper left.  The angular scale is shown by
  the bar at the bottom left.  The inset shows the lens system with no labels overlaid.
  }
\label{fig:color_picture}
\end{centering}
\end{figure*}

In this section, we describe the photometric and spectroscopic
properties of the system.  The lens is located at (R.A., Dec.)=($14^h24^m49\mbox{\ensuremath{.\!\!^{\mathrm{s}}}}0$, $-00^\circ53^{\prime}21\farcs65$)
(J2000).  A multicolor image of \horus~constructed from HSC imaging data is
shown in Figure~\ref{fig:color_picture}.   The two sources with distinct colors are clearly
seen around the lens: a red inner source (S1) and a blue Einstein ring (S2). These arcs are
approximately 3\arcsec~in radius.  Several knots with similar colors are embedded in the arcs,
which are useful features to constrain our lens models (Section~\ref{sec:lensmod}).

\subsection{Photometric Data}
\horus~is located in the Wide layer of the HSC survey, which aims to cover
$\sim$1400 deg$^{2}$ down to $r\sim26$.  This paper is based on the S15A internal data release.
The data are processed with hscPipe v3.8.5, a version of Large Synoptic Survey Telescope pipeline
\citep{ivezic2008,axelrod2010} calibrated against PanSTARRS1 \citep{schlafly2012,tonry2012,magnier2013}.
A summary of the HSC data on this system is given in Table \ref{tab:phot}.
The exposure times are short, but reach fairly deep thanks to the superb image quality.
We find that the system has also been observed by KiDS \citep{dejong2013} and VISTA-VIKING \citep{edge2013}.
However, these shallower data do not reveal the double-source nature of
the system -- only the outer arc is seen in the KiDS data, and VISTA-VIKING
detect only the few brightest knots.  We use only the photometric data from HSC in this paper.

\begin{table}
  \begin{center}
    \begin{tabular}{cccc}
      \hline
      filter & seeing (\arcsec) & exposure time (s)\tablenotemark{a} & $5\sigma$ depth\tablenotemark{b}\\
      \hline
      $g$    & 0.81            & 450    &  26.7\\
      $r$    & 0.62            & 450    &  26.4\\
      $i$    & 0.61            & 1600   &  26.4\\
      $z$    & 0.66            & 1600   &  25.5\\
      $y$    & 0.66            & 2000   &  24.6\\
      \hline
    \end{tabular}
  \caption{Photometric data.}
  \tablenotetext{1}{Some of the $y$-band data are taken under poor conditions.}
  \tablenotetext{2}{$5\sigma$ depth for point sources.  The depths within 1\farcs5 apertures, which may be more relevant for extended sources, are shallower by $\sim 0.2$ mag.}
  \label{tab:phot}
  \end{center}
\end{table}

\subsection{Spectroscopic Data}
The lens galaxy has a spectroscopic redshift of $z_{\mathrm{L}}=\zl$ from the Sloan Digital
Sky Survey \citep{alam2015} and we discuss its properties in Section~\ref{sec:lensgal}.
In order to determine the redshifts of the background sources, we performed follow-up spectroscopic
observations with the Folded-port Infrared Echellette \citep[FIRE;][]{simcoe2013} on
the 6.5-m Magellan I Baade telescope.  FIRE covers the wavelength range $0.82-2.51~\mu$m
at a spectral resolution of $R=6000$.  The observations were taken on the nights of
UT 2016 February 15--16.  Conditions were clear, and the seeing varied from $0\farcs6-1\farcs0$.
A slit width of 1\arcsec~was used to image the primary features of S1 and S2 across six
separate pointings with exposure times ranging from $10-40$ minutes.  The data are reduced
using a combination of the FIREHOSE pipeline and custom IDL scripts.

\begin{figure*}[h]
\begin{centering}
\epsscale{1.1}
\plotone{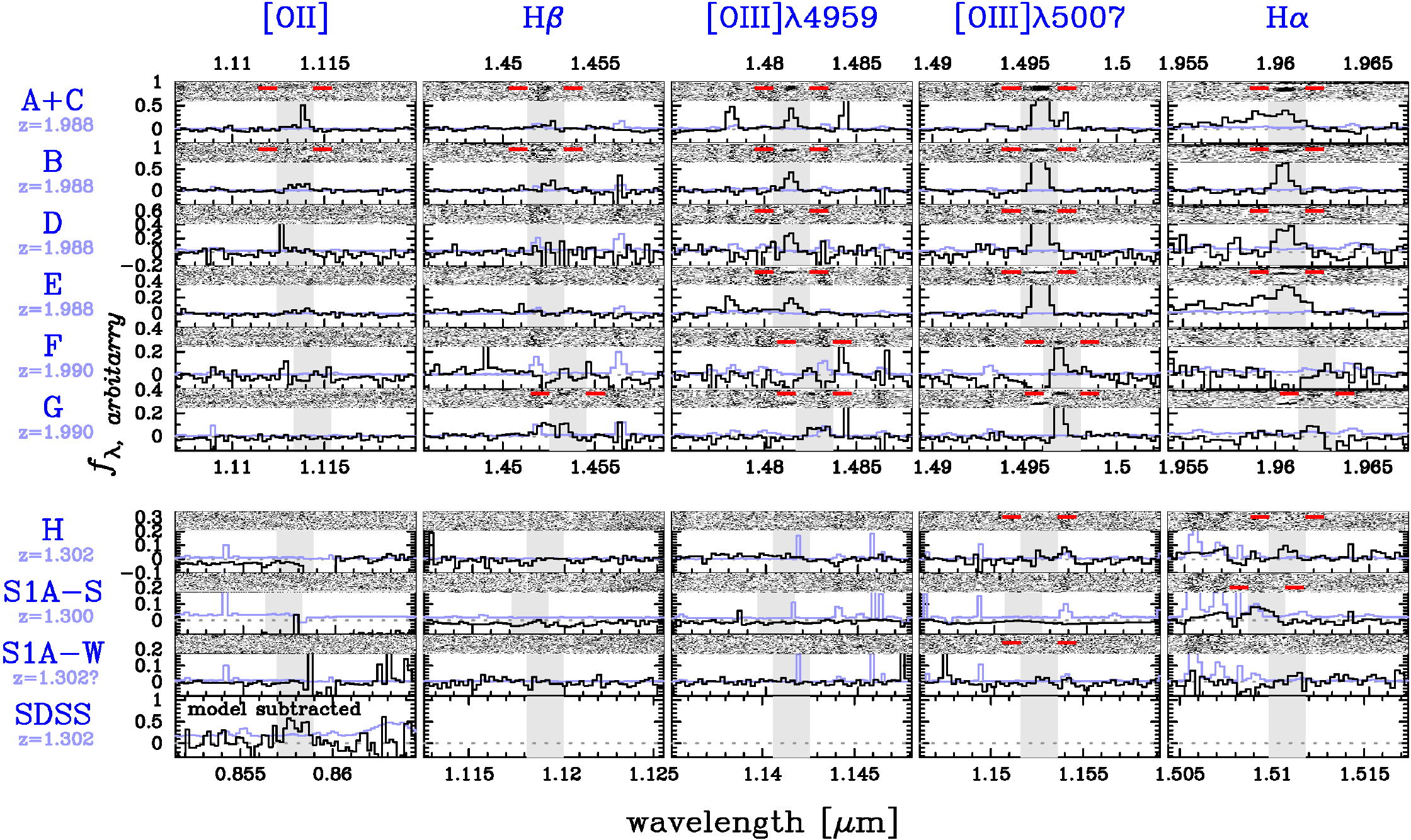}
\caption{
  1D and 2D spectra of the arcs and knots taken with FIRE.  The black and light-blue spectra are
  the 1D object and noise spectra, respectively, and are binned over $2~\rm \AA$.
  The shaded regions show where strong emission lines are located.  Not all the lines
  are securely detected and we mark in red in the 2D spectra the secure or likely line detections.
  Note the small redshift offset of F and G with respect to A--E.  There is also a small offset
  between H and S1A-S.  The bottom panel shows the spectrum of the lens from SDSS with the best-fit
  continuum spectrum subtracted. The [OII] emission from H is observed (see text for details).
}
\label{fig:spectra}
\end{centering}
\end{figure*}

We successfully determine the redshifts of most of the main features of both sources indicated
in Figure~\ref{fig:color_picture}, making this the first DSP lens with
  spectroscopic confirmation of both source redshifts.  As shown in Figure~\ref{fig:spectra},
we measure the redshift of the outer source to be $z_{\mathrm{S}2}=\zstwo$ based on multiple detections
of the {\sc [oii]}, H$\beta$, {\sc [oiii]} and H$\alpha$ lines.  Interestingly, S2 has two components that
are slightly offset by $\Delta z \approx 0.002$ in redshift space.  However, they are likely physically
associated given the small difference.

The inner source is confirmed at $z_{\mathrm{S}1}=\zsone$ with 
weaker detections of the {\sc [oiii]} and H$\alpha$ lines.  In addition to our FIRE spectra,
the spectrum from SDSS also confirms the redshift of S1 -- feature H is included within the SDSS fiber
targetting the lens galaxy and we observe clear {\sc [oii]} emission from H once the continuum 
is subtracted.  S1 also seems to have two components as indicated by the small redshift
offset of $\Delta z \approx 0.002$ between H and the southern segment of the arc (S1A-S) as well as a very marginal hint of
the western segment (S1A-W) being at a redshift consistent with H, suggesting that S1A-W is the counterimage of H.
Further follow-up observations will reveal the detailed dynamics of each source.

\section{Lens Galaxy Properties} \label{sec:lensgal}
The lens galaxy is a passive galaxy with no clear signs of active star formation based on the SDSS spectrum.
Preliminary results of the {\sc CAMIRA} algorithm \citep{oguri2014} applied to the HSC survey data suggest that there is a rich cluster of galaxies at $z_{\mathrm{phot}} \sim 0.81$ at the location of the lens.  The richness of $N\sim50$ implies a cluster mass of $M_{\mathrm{vir}} \sim 7\times10^{14} M_{\odot}$.  Given the lens galaxy's large stellar
mass (estimated below) and proximity to the cluster center, it is potentially the brightest cluster galaxy.  A more detailed analysis of the lens environment based on multi-object spectroscopy from the Inamori Magellan Aerial Camera and Spectrograph \citep[IMACS;][]{dressler2011} will be presented in a follow-up work (Wong et al. in preparation).

\begin{figure*}[ht]
\begin{centering}
\includegraphics[width=0.33\textwidth]{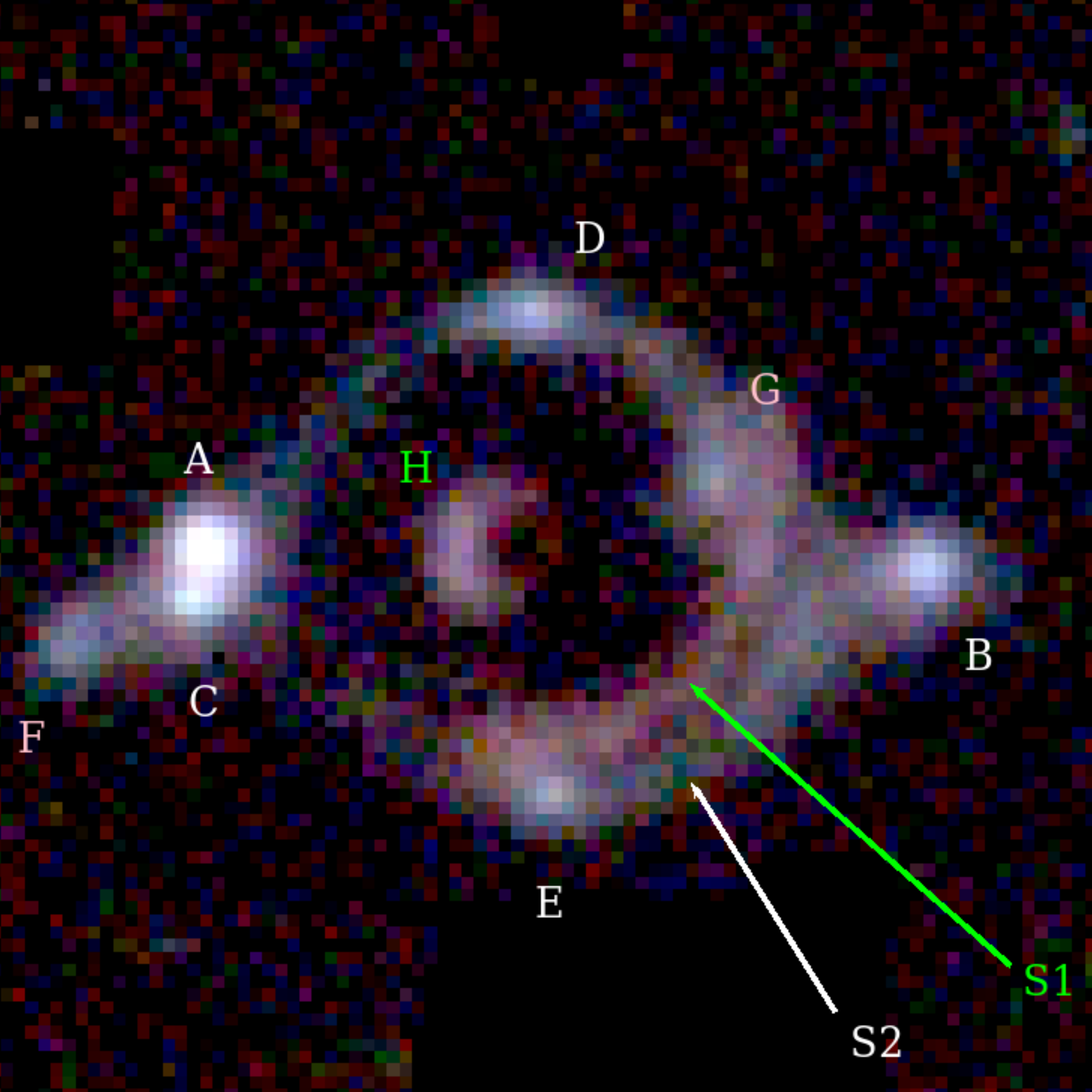}
\includegraphics[width=0.33\textwidth]{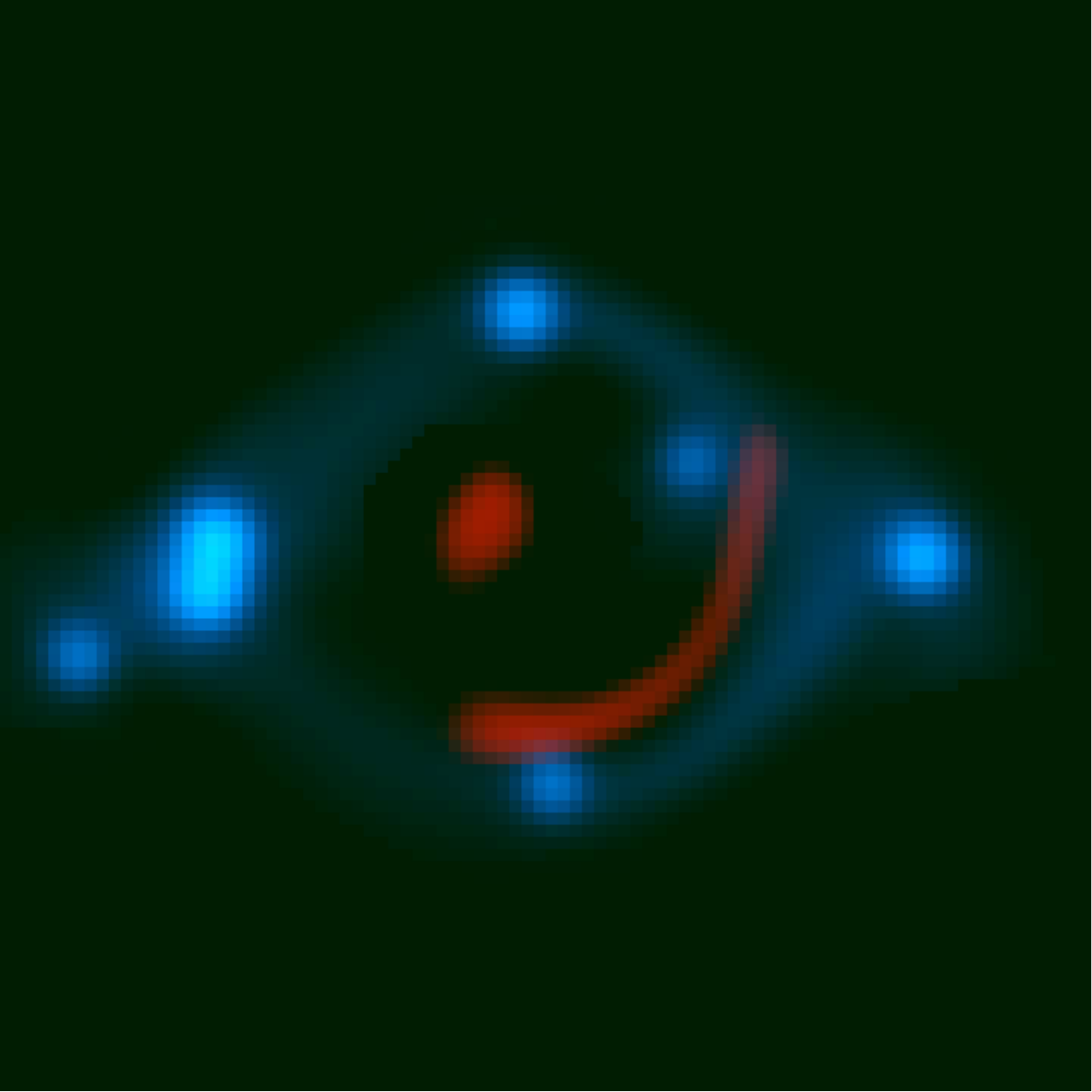}
\includegraphics[width=0.33\textwidth]{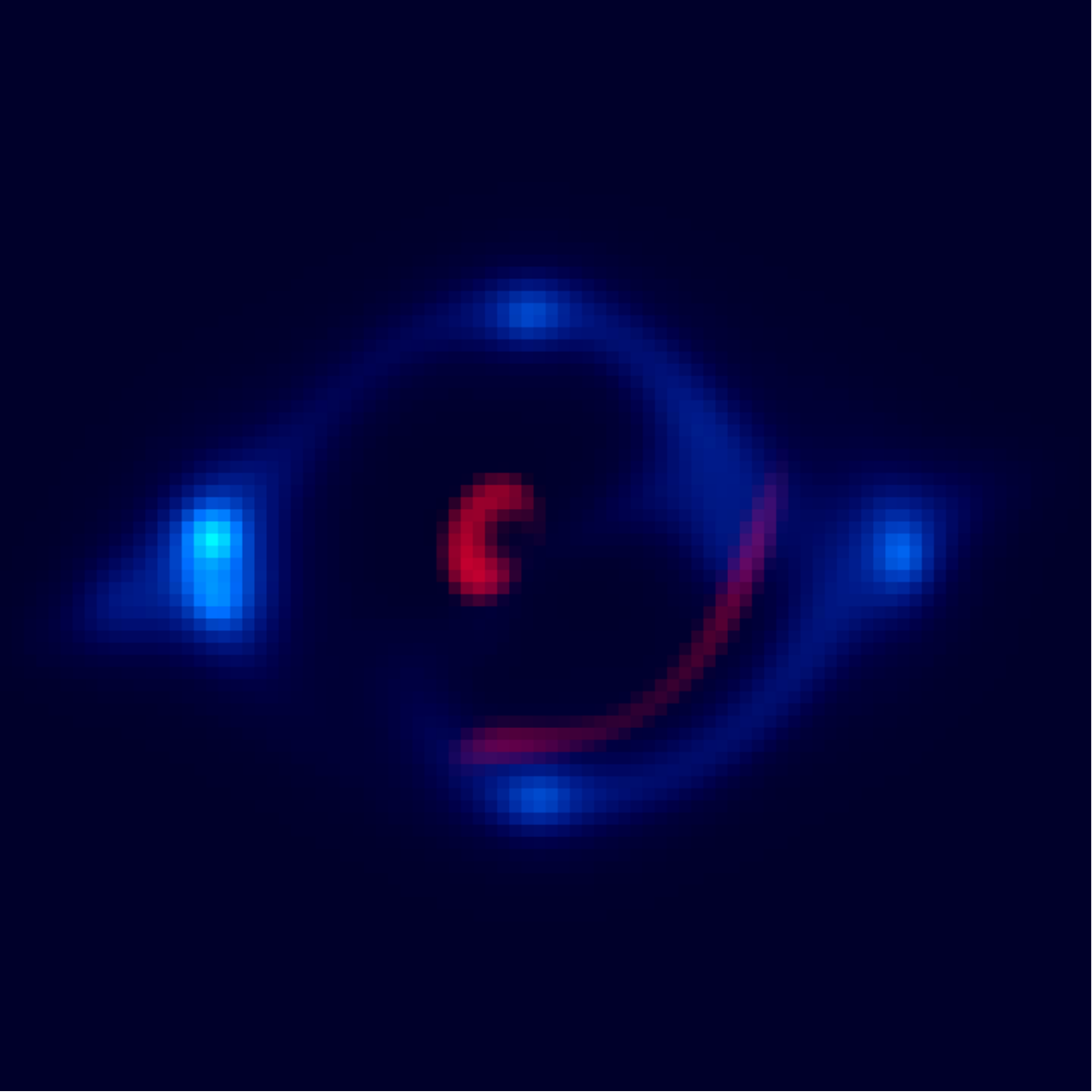}
\caption{
  {\bf Left:}
  Lens-subtracted $riz$ image. Feature H is clearly seen once the lens galaxy is subtracted.
  {\bf Center:}
  Reconstructed image configuration from {\sc Glee}.  The red source represents S1, while the blue source represents S2.
  {\bf Right:}
  Same as center panel, but from {\sc Glafic}.  Both models require an additional mass component to split features A and C into two distinct components.
}
\label{fig:lenssub}
\end{centering}
\end{figure*}

Despite the superb seeing, the lens and the surrounding arcs are still blended in the HSC imaging.
In order to study the lens and perform detailed analyses of the background sources, we fit
the lens with a double S\'{e}rsic model and subtract it from the images
using {\sc Glee} \citep{suyu2010,suyu2012}.  We first fit the lens in the $i$-band, in which
the lens is detected at a high S/N of $\sim500$ within a $1\farcs5$ aperture and
the seeing is good ($0\farcs61$; see Table~\ref{tab:phot}).
The PSF is modeled in each exposure and is evaluated at the position of the lens
by coadding the PSF models.
We then fit the lens in the other bands
with the structural parameters such as centroid, effective radius, ellipticity, 
and position angle fixed to those measured in the $i$-band.
Only the flux normalization is allowed to vary.  The PSF model in each band is
evaluated in the same way as the $i$-band.  The multicolor lens-subtracted image 
is shown in the left panel of Figure~\ref{fig:lenssub}. An arc-like feature (H) on the left of the lens emerges.
The images show no
subtraction residuals at the location of the lens, demonstrating our
accurate modeling of PSF and the lens galaxy light.

Using the 5-band photomtetry and the known redshift of the lens
galaxy, we fit its spectral energy distribution (SED) using {\sc mizuki}
\citep{tanaka2015}.
Due to the strong blending, we use the double S\'{e}rsic fit to
the lens to estimate the flux in each band.  The fit has a very extended
envelope out to large radii, but we truncate the model at a radius of
$3\farcs3$ ($\sim25$ kpc at the lens redshift), which is the radius of the S2 arc.
This is a somewhat arbitrary choice, but it should include most of the lens light
and also represent the enclosed stellar mass while keeping the flux from
the (uncertain) envelope relatively small.
The SED is then fit with a suite of model templates generated by
\citet{bruzual2003} models assuming a \citet{chabrier2003} IMF and \citet{calzetti2000} dust attenuation curve.
We apply Bayesian priors on the physical properties of galaxies to keep
the model parameters within realistic ranges and to effectively let the templates evolve with redshift in an observationally motivated way \citep[for more details, see][]{tanaka2015}.
We find that the lens galaxy is very massive -- it has a stellar mass of
$6.6^{+0.7}_{-0.1}\times10^{11}~\rm M_\odot$, making it among the most massive galaxies around
this redshift \citep{muzzin2013}.  The star formation rate from the fit is
very small for its stellar mass (SFR $=0.9^{+1.3}_{-0.1}~\rm M_\odot~yr^{-1}$), as is the dust attenuation ($\tau_V=0.2^{+0.2}_{-0.1}$). Together with the early-type morphology, the lens
may be the brightest cluster galaxy.  Note that the given uncertainties are just
statistical uncertainties and do not account for systematic effects.

\section{Lens Models}  \label{sec:lensmod}
We model the system based on the data in hand using two independent codes:
{\sc Glee} \citep{suyu2010,suyu2012} and {\sc Glafic} \citep{oguri2010}.
The system is complex, and our goal
is to reproduce the main features of the image configuration of both sources to obtain
a basic understanding of the system.  More detailed lens models will be presented
in a future paper (Wong et al. in preparation).

\subsection{Pixelated Source Modeling}
We first model \horus~with {\sc Glee} using pixelated grids with curvature regularization for the extended sources
\citep{suyu2006}.
Lensing mass distributions are described by parameterized profiles.  Model parameters of the lens and source are constrained through Markov Chain Monte Carlo (MCMC) sampling.  We use the full multiplane lens equation \citep[e.g.,][]{blandford1986} to calculate deflection angles since there are multiple lens and source planes involved (Suyu et al. in preparation).

Our modeling strategy is to first use only S1 to constrain the parameters of the main lens galaxy (L), then to model L and S1 simultaneously with the light from S2 as constraints.  We separate S1 and S2 by subtracting the HSC images in two filters (after matching the seeing) weighted such that the light from either S1 or S2 is removed and leaving only light from the other.  We find that using the $i$ and $z$ bands give the cleanest separation, as these filters straddle the 4000 \AA~break of S1 and probe the UV continuum of S2.  The lens galaxy L is parameterized as a singular power-law elliptical mass distribution \citep{barkana1998}.  We fix the centroid to the galaxy light centroid and adopt flat priors on the ellipticity, orientation, and mass profile slope.  We do not include external shear, as it does not drastically improve the fit.

For the second part of our modeling, S1 is parameterized as a singular isothermal ellipsoid (SIE) with flat priors on its ellipticity and orientation.  The centroid of S1 is given a flat prior, but we link the centroid to the weighted source position of the brightest pixels in H and S1A.  The likelihood determined from the offset between these pixels and the predicted image positions of the S1 centroid (assuming one HSC-pixel positional uncertainty of $\sim 0\farcs17$) are factored into the total likelihood.  We then simultaneously vary the parameters of the lens and S1 using the light distribution of S2 as constraints.  Using a pixelated source alone is unable to remove point-like residuals at the positions of the bright clumps in S2, so we also fit point sources convolved with the PSF at each of these seven locations (A--G in Figure~\ref{fig:color_picture}), allowing their position and amplitude to vary.  We also include a satellite galaxy (L1) parameterized as a singular isothermal sphere (SIS) at the main lens redshift between features A and C, as a smooth model is unable to account for these features.  L1 is given a flat prior on its position and Einstein radius.

\subsection{Parameterized Source Modeling}
{\sc Glafic} uses parameterized models for the extended light profile of
the lensed source, contrary to 
{\sc Glee}. While the current public version of {\sc Glafic} supports a single lens plane only, we recently implemented the ability to calculate the deflection
angles over two lens planes using the standard multiplane lens equation formalism, which is essential for modeling of this system.

In principle, {\sc Glafic} can fit the light and mass profile of
the lens galaxies simultaneously. However, for simplicity, we model the light
profile of the main lens galaxy (L) first with a S\'{e}rsic profile and subtract
it from the $i$-band image. While holding the lens position fixed from the
S\'{e}rsic model, we choose an elliptical power-law to model the lens potential.  

We adopt the following strategy for modeling each lensed source. A
single S\'{e}rsic model is used to describe the light profile of S1.  The
lens model includes an elliptical power-law for galaxy L and an SIS for
the dark satellite L1, assumed to be at the same redshift as L.
For S2, we need two-component S\'{e}rsic models to fit the compact
(knots A,B,C,D and E) and extended (blue ring) features. We also need
another S\'{e}rsic component to model the knots G+F, which 
correspond to a distinct source (see
Figure~\ref{fig:spectra}).  For S2, the mass model is same as before (L+L1), 
except now there is a secondary lens plane to account for the
mass of S1, which is modeled as an SIE. The center of
the light profile of source S1 is the same as the associated mass
component. 

Both sources are modeled simultaneously but the likelihood of the lens
model for S1 is calculated by masking out everything else in the
image except H and S1A.  Similarly, the likelihood of the lens model for
S2 is calculated by masking out everything except the blue features.
The final likelihood is a product of the two likelihoods. We use {\sc Glafic} for
solving the lens equations while the sampling of the posterior
distribution of the model parameters is done via the Python module {\it
emcee} \citep{foremanmackey2013}.

\section{Results and Summary}  \label{sec:summary}
The reconstructed image configurations are shown in the middle and right panels of Figure~\ref{fig:lenssub}.
We find that both {\sc Glee} and {\sc Glafic} successfully reproduce the main features of the system.
Both methods prefer a shallow mass profile ($\gamma \sim 1.5$, where $\rho\propto r^{-\gamma}$) for L,
suggesting that there is a large dark matter fraction, possibly due to the cluster environment.
Both models also require an additional mass component between images A and C to split this feature into two images.
We have assumed that this mass component, which is undetected in the HSC imaging and FIRE spectroscopy,
is at the same redshift as the lens galaxy, although it could potentially be a line-of-sight structure as well.
The image splitting is unlikely to be due to S1 as the positions are not coincident.
The velocity dispersion of L1, assuming an SIS profile, is $\sim 60-100$ km s$^{-1}$, although the uncertainties
are large and there are likely degeneracies with the centroid position relative to features A and C.
Deeper and higher-resolution imaging is needed to potentially detect L1.

Both models find that S1 has a velocity dispersion of $\sim 230$ km s$^{-1}$ (assuming an SIE profile) 
with moderate to high ellipticity.
The {\sc Glee} model maps feature H to S1A-W, while S1A-S maps to a segment of H just above the lens galaxy, resulting in a source that either consists of two components or is highly elongated (e.g. an edge-on disk).  The slight velocity offset of H and S1A-W with S1A-S indicated by the FIRE data supports this interpretation.  Unfortunately, the upper segment of H was not targeted in our FIRE observations.  There are likely degeneracies among the physical properties of L and S1.  More detailed mass models using improved data will explore these degeneracies and more accurately determine the model parameters.

By extrapolating the results from CFHTLS \citep[i.e., one confirmed DSP lens from][]{tu2009}, we expect to discover $\sim 10$ DSP lenses in the HSC survey, of which \horus~is the first.  Our initial results highlight the potential wealth of information provided by DSP lenses, as well as the challenges of DSP lens modeling and the importance of high-quality data and multiple lens model analyses to understand all of the degeneracies and systematic effects.  The spectroscopic confirmation of both sources and the presence of the potential substructure causing the image splitting of features A and C make \horus~a valuable system for studies of cosmology and galaxy structure.


\acknowledgements
We thank Aleksi Halkola for his contributions to this project.
This work is based on data collected at Subaru Telescope, which is operated by the National Astronomical
Observatory of Japan, and is supported by JSPS KAKENHI Grant Numbers JP15K17617, JP26800093, and JP15H05892.
SHS gratefully acknowledges support from the Max Planck Society through the Max Planck Research Group.
We thank the referee for a helpful report.
The Hyper Suprime-Cam (HSC) collaboration includes the astronomical
communities of Japan and Taiwan, and Princeton University.  The HSC
instrumentation and software were developed by the National
Astronomical Observatory of Japan (NAOJ), the Kavli Institute for the
Physics and Mathematics of the Universe (Kavli IPMU), the University
of Tokyo, the High Energy Accelerator Research Organization (KEK), the
Academia Sinica Institute for Astronomy and Astrophysics in Taiwan
(ASIAA), and Princeton University.  Funding was contributed by the FIRST 
program from Japanese Cabinet Office, the Ministry of Education, Culture, 
Sports, Science and Technology (MEXT), the Japan Society for the 
Promotion of Science (JSPS),  Japan Science and Technology Agency 
(JST),  the Toray Science  Foundation, NAOJ, Kavli IPMU, KEK, ASIAA,  
and Princeton University.
This paper makes use of software developed for the Large Synoptic Survey Telescope.
We thank the LSST Project for making their code freely available\footnote{http://dm.lsstcorp.org}.
The Pan-STARRS1 (PS1) Surveys  have been made possible through contributions of the Institute for
Astronomy, the University of Hawaii, the Pan-STARRS Project Office, the Max-Planck Society and
its participating institutes, the Max Planck Institute for Astronomy and the Max Planck
Institute for Extraterrestrial Physics, The Johns Hopkins University, Durham University,
the University of Edinburgh, Queen's University Belfast, the Harvard-Smithsonian Center for Astrophysics,
the Las Cumbres Observatory Global Telescope Network Incorporated, the National Central University of Taiwan,
the Space Telescope Science Institute, the National Aeronautics and Space Administration under Grant
No. NNX08AR22G issued through the Planetary Science Division of the NASA Science Mission Directorate,
the National Science Foundation under Grant No. AST-1238877, the University of Maryland, and Eotvos
Lorand University (ELTE).


\end{document}